\documentclass[conference]{IEEEtran}
\IEEEoverridecommandlockouts

\usepackage{cite}
\usepackage{amsmath,amssymb,amsfonts}
\usepackage{algorithmic}
\usepackage{graphicx}
\usepackage{textcomp}
\usepackage{xcolor}
\usepackage{balance}
\usepackage[hyphens]{url}

\usepackage{subcaption}
\usepackage{quantikz}
\usepackage{hyperref}
\usepackage{booktabs}

\def\BibTeX{{\rm B\kern-.05em{\sc i\kern-.025em b}\kern-.08em
    T\kern-.1667em\lower.7ex\hbox{E}\kern-.125emX}}
\begin{document}

\pdfpagewidth=8.5in
\pdfpageheight=11in

\newcommand{\iscasubmissionnumber}{NaN}

\pagenumbering{arabic}

\title{A Scalable Open-Source QEC System with Sub-Microsecond Decoding-Feedback Latency}

\author{\IEEEauthorblockN{Junyi Liu\IEEEauthorrefmark{1}, Yi Lee\IEEEauthorrefmark{1}, Yilun Xu\IEEEauthorrefmark{2}, Gang Huang\IEEEauthorrefmark{2}, and Xiaodi Wu\IEEEauthorrefmark{1}}
\IEEEauthorblockA{\IEEEauthorrefmark{1}Joint Center for Quantum Information and Computer Science, University of Maryland, College Park, MD} 
\IEEEauthorblockA{\IEEEauthorrefmark{2}Lawrence Berkeley National Laboratory, Berkeley, CA} 
\href{mailto:xiaodiwu@umd.edu}{xiaodiwu@umd.edu}
\vspace{-7mm}
}

\maketitle
\thispagestyle{plain}
\pagestyle{plain}


\begin{abstract}
Quantum error correction (QEC) is essential for realizing large-scale, fault-tolerant quantum computation, yet its practical implementation remains a major engineering challenge. In particular, QEC demands precise real-time control of a large number of qubits and low-latency, high-throughput and accurate decoding of error syndromes. While most prior work has focused primarily on decoder design, the overall performance of any QEC system depends critically on all its subsystems including control, communication, and decoding, as well as their integration.

To address this challenge, we present an open-source\footnote{\url{https://github.com/Wu-Quantum-Application-System-Group/RISC-Q}}, fully integrated QEC system built on RISC-Q, a generator for RISC-V–based quantum control architectures. Implemented on RFSoC FPGAs, our system prototype integrates real-time qubit control, a scalable distributed multi-board architecture, and the state-of-the-art hardware QEC decoder within a low-latency, high-throughput decoding pipeline, forming a complete hardware platform ready for deployment with superconducting qubits.

Experimental evaluation on a three-board prototype based on AMD ZCU216 RFSoCs demonstrates an end-to-end QEC decoding-feedback latency of 446 ns for a distance-3 surface code, including syndrome aggregation, network communication, syndrome decoding, and error distribution. 
Extrapolating from measured subsystem performance and state-of-the-art decoder benchmarks, the architecture can achieve sub-microsecond decoding-feedback latency up to a distance-21 surface code ($\sim$881 physical qubits) when scaled to larger hardware configurations.
\end{abstract}

\section{Introduction}
Quantum computing promises to solve problems that are classically intractable, but doing so will require performing vast numbers of operations across systems containing thousands of qubits \cite{gidney2021factor}.
However, quantum systems are inherently fragile: their error rates are several orders of magnitude higher than those of classical digital systems~\cite{google2024qec,schroeder2009dram,kim2014flipping}.
Consequently, quantum error correction (QEC)~\cite{RevModPhys.87.307} is an essential prerequisite for achieving any practically useful quantum computation.

Rapid and significant experimental advances in quantum error correction (QEC) have emerged across diverse hardware platforms, propelled by intense efforts from both academia  (e.g.,~\cite{bluvstein2024logical,qdit-break-even}) and industry (e.g.,~\cite{google2024qec,QuERA-magic,reichardt2025faulttolerantquantumcomputationneutral,yamamoto2025quantumerrorcorrectedcomputationmolecular,aws-cat-qubit,caune2024demonstrating}), underscoring its foundational importance for scalable quantum computing.

However, a scalable QEC system would require precise real-time control over large numbers of qubits, together with low-latency, high-throughput, and accurate syndrome decoding and feedback. Realizing large-scale, fault-tolerant quantum computation is therefore not only a challenge in advancing quantum hardware, but also a substantial engineering undertaking in developing the accompanying classical control and error-correction infrastructure~\cite{mohseni2025buildquantumsupercomputerscaling, battistel2023real}.

More precisely, the desirable performance of any real-time QEC system can be characterized by four key metrics~\cite{battistel2023real,google2024qec,caune2024demonstrating}: \emph{accuracy, throughput, latency}, and \emph{scalability}:
\begin{itemize}
  \item \textbf{Accuracy} quantifies the probability that the decoder correctly infers the underlying errors from measured syndromes and also reflects the precision of (synchronized) control across many qubits. 
  \item \textbf{Throughput} refers to the volume of syndrome data processed per unit time, encompassing both the communication bandwidth for transporting syndromes and the decoder’s processing capacity. 
  \item \textbf{Latency} is the total time from acquiring syndrome data to applying the corresponding feedback control, including all  system overheads, i.e. \emph{decoding-feedback} latency. It directly limits the execution speed of non-Clifford gates and, in turn, the overall clock rate of a quantum computer.
  \item \textbf{Scalability} refers to the capability of a system architecture to maintains high accuracy, throughput, and low latency as it scales to large numbers of qubits, without incurring excessive resource or design overhead.
\end{itemize}

Superconducting qubit systems arguably impose the most stringent requirements on QEC systems to date, demanding microsecond-scale latency and gigabits-per-second throughput for quantum processors with up to thousands of qubits. 
While most prior work (e.g.,~\cite{das2022afs,das2022lilliput,vittal2023astrea,liao2023wit,liyanage2024fpga,barber2025real,wu2025micro,liyanage2025network,maan2025decoding,maurer2025real,zhang2025latte}) has focused on optimizing decoder performance---a critical yet partial contributor of the overall system performance---a holistic design encompassing the entire QEC control stack is essential.

While conceptual QEC architectures have been proposed \cite{byun2022xqsim,kim2024fault}, 
the only publicly known integrated QEC systems are developed by Google~\cite{google2024qec} and by Rigetti-Riverlane~\cite{caune2024demonstrating}. 
For real-time decoding, Google reports a 63 \textmu s decoding latency for a distance-5 surface
code with 10 syndrome measurements on 49 qubits, an simulated 108 \textmu s decoding latency for a distance-7 surface code,  
and a roughly 10 \textmu s communication latency to transport syndrome measurements to the decoder. 
R\&R reports a 6.5 \textmu s decode-and-feedback latency with 9 syndrome measurements and a $\sim$2 \textmu s communication latency for an 8-qubit stability protocol. 
Both systems leave substantial room for improvement in terms of latency and scalability.
In particular, a central open question is whether end-to-end latency can be reduced below the 1.1 \textmu s
QEC cycle time of Google's machine~\cite{google2024qec} while scaling up to hundreds of qubits.

In this work, we design and implement a full-stack, integrated quantum error-correction control system that meets these stringent requirements through a comprehensive co-design of system architecture, hardware, and communication infrastructure. Specifically, our system has following features: 
\begin{itemize}
    \item \emph{Fully Hardware Integration}: We implement the entire control stack (e.g., pulse generation, synchronization) and the full QEC pipeline (e.g., syndrome processing, decoding, and feedback) directly in hardware, integrating all modules through high-bandwidth, low-latency data paths that maximize parallelism and eliminate the overhead and timing uncertainty inherent to software-based systems.
    \item \emph{Scalable Multi-Core/Board Design:} To ensure scalability and high throughput, we adopt a distributed control architecture where each qubit is managed by its dedicated control core. Tight synchronization and efficient inter-core communication are built into the design, enabling large-scale parallel operation while preserving timing accuracy and system coherence as the qubit count grows.
    \item \emph{Modular-design and Interoperability: } Our QEC system employs a modular architecture, enabling independent development and seamless interoperability of control cores, decoders, and communication modules. This design supports predictable scaling, allowing latency and throughput to be confidently extrapolated as the system grows.
    \item \emph{Flexible Programmability:} To program our QEC system, the software stack adopts a clean separation between the configuration logic (Python/PYNQ layer) and the real-time execution (C/RISC-V layer), offering high programmability and orthogonal configurability while preserving deterministic, hardware-level execution. This co-design enables a practical and scalable development environment for distributed quantum control.
\end{itemize}

Built on RISC-Q, a generator~\cite{liu2025risc} for RISC-V–based quantum control architectures, our system is fully open-source and RISC-V–compatible. This enables efficient development and testing of the on-board QEC system by leveraging the rich RISC-V toolchain.
Our prototype is implemented on three AMD ZCU216 RFSoC boards and the experimentally validated QubiC hardware system~\cite{xu2023qubic}, making it readily deployable for controlling real superconducting qubits, with each leaf node running at 500 MHz and supporting up to 14 qubits.
A direct measure of our system reveals a total QEC decoding-feedback
latency of 446 ns for a distance-3 surface code with 3 syndrome measurement rounds.
Leveraging our measured data and modular design, we can reliably extrapolate the system's performance for larger surface codes: our analysis indicates that the proposed system can scale to distance 21 ($\sim$881 physical qubits) with decoding-feedback latency below 1 \textmu s, confirming the feasibility of Google's 1.1 \textmu s target. 

In summary, we make the following contributions: 
\begin{enumerate}
    \item We design a scalable, modular, distributed multi-board real-time QEC system that fully integrates control and the entire QEC pipeline in hardware, interconnected through high-bandwidth, low-latency data paths.
    \item We implement a fully open-source, RISC-V–compatible prototype that is readily deployable to superconducting qubit systems, with measured performance indicating sub-microsecond decoding–feedback latency up to a distance-21 surface code ($\sim$881 physical qubits). 
\end{enumerate}

\section{Background}
This section presents an overview of the qubit control architecture and quantum error correction operations, emphasizing the timing and performance requirements necessary for the integrated system to achieve precise real-time control of a large number of qubits, as well as low-latency, high-throughput, and accurate decoding of error syndromes.

\subsection{Qubit Control System}
Modern quantum computing platforms---such as superconducting circuits, trapped ions, and neutral atoms---depend on precisely engineered RF signals to manipulate quantum states. Despite differences in physical implementation, they share a fundamental need for accurately timed, phase-coherent control of sensitive, short-lived quantum systems. Unlike classical control, quantum control must maintain precise phase coherence at high frequencies and operate within the strict latency limits imposed by qubit coherence times \cite{google2024qec,caune2024demonstrating}. 

Superconducting qubits, among the most demanding physical platforms, typically require control signals in the 3–6~GHz range. The phase of these RF signals determines the rotation axis of a quantum gate, while the amplitude and envelope define the rotation angle. At this frequency, even a single-cycle timing error can lead to a measurable loss in gate fidelity~\cite{krantz2019quantum}. Maintaining phase coherence across multi-channel gigahertz controls is a key engineering challenge as qubit counts grow. In addition, the short coherence time of superconducting qubits, on the order of tens of microseconds, requires that measurement decoding and feedback control be completed within microsecond-scale latency windows~\cite{google2024qec,caune2024demonstrating}.
A complete qubit control system typically comprises three primary components:
\begin{itemize}
  \item an \emph{RF signal generator}, which produces precisely shaped analog RF control signals to drive qubit operations;
  \item a \emph{readout decoder}, which extracts digital measurement outcomes from analog detector signals; and
  \item a \emph{controller}, which orchestrates pulse sequences and feedback logic in real time.
\end{itemize}

\subsubsection{RF Signal Generator}
The modern approach for RF signal generation is Direct Digital Synthesis (DDS), which digitally computes discrete waveform samples and converts them to analog signals through a Digital-to-Analog Converter (DAC). In contrast to traditional AWG–mixer architectures that upconvert baseband signals to the RF domain, DDS enables direct digital control of frequency and phase, offering faster reconfiguration, and reduced calibration complexity \cite{stefanazzi2022qick,xu2023qubic}. 

Among all components of the control system, the RF generator imposes the most stringent real-time requirements. Because the phase of the carrier directly determines the gate axis, even a single clock-cycle offset at gigahertz frequencies (corresponding to sub-nanosecond errors) can cause a substantial phase deviation. To prevent such errors, the generator must operate with cycle-level determinism and be synchronized to a global reference clock shared across all control channels. This ensures that phase coherence is preserved even when multiple gates are executed concurrently across different qubits.

\subsubsection{Readout Decoder}
The readout decoder converts analog measurement signals into discrete binary measurement outcomes (0 or 1). The nature of the readout signal varies depending on the quantum hardware platform: in trapped-ion or neutral-atom systems, the signal is typically optical, captured as photon counts or image pixels; in superconducting systems, it is an RF signal, where the qubit state is inferred from the phase of a reflected RF pulse.

For real-time feedback applications such as quantum error correction, both latency and accuracy of the decoding process are critical. The decoding must be performed fast enough to feed results back into subsequent control steps without exceeding the microsecond-level timing budget. At the same time, the readout process must achieve high classification accuracy despite the presence of amplifier and thermal noise, which otherwise contribute to error accumulation.

\subsubsection{Controller}
The controller serves as the programmable interface between, e.g., quantum algorithms, error correction protocols, and the underlying control hardware.
It must support flexible sequencing of gate operations, conditional branching based on mid-circuit measurements, and integration with external timing and synchronization signals.
Such flexibility is typically achieved using general-purpose processors, allowing pulse programs and feedback logic to be defined in software rather than fixed hardware.

Beyond programmability, the controller must also meet strict real-time constraints to coordinate all components.
It must interface with heterogeneous peripherals, such as RF generators, readout decoders, and memory subsystems, through low-latency communication buses.
Clock frequencies of several hundred megahertz are typically required to match the RF signal update rate, and overall response time must remain below 1 \textmu s to support real-time QEC.
These requirements represent key challenges in designing controllers for scalable quantum-control architectures.

\subsection{Quantum Error Correction Operations}
Quantum error correction preserves fragile quantum information by encoding a logical qubit into multiple physical qubits. Errors are identified and corrected via repeated stabilizer measurements, which yield syndromes that locate errors without collapsing the encoded state.
Given the much higher error rates of physical qubits than classical 
~\cite{google2024qec,schroeder2009dram,kim2014flipping}, QEC is essential for realizing large-scale quantum computations.

\subsubsection{Syndrome Measurement}
The surface code~\cite{fowler2012surface} is a leading type of stabilizer codes~\cite{gottesman1997stabilizer}, where errors syndromes are measured from entangled qubits. In a stabilizer code, each stabilizer corresponds to a parity check of errors across a set of qubits. The measurement outcome of a stabilizer, or syndrome, is the parity value, analogous to parity bits in classical codes. 

Directly measuring each data qubit would collapse the logical state. Instead, an ancilla qubit is used to measure the stabilizer indirectly. Thus, in a typical stabilizer code, each syndrome measurement round involves two steps:
\begin{itemize}
  \item Entangling data qubits with auxiliary ancilla qubits, extracting error information onto the ancillas; and
  \item Measuring the ancillas to extract binary syndromes that encode the parity of error events.
\end{itemize}

At the hardware level, each entangling gate is realized by a specific pulse. In superconducting systems, qubits’ sensitivity to timing and phase requires pulses at gigahertz frequencies with precise phase coherence, necessitating real-time RF generation via high-speed DACs under strict timing control. In the same system, qubit readout uses dispersive measurement~\cite{krantz2019quantum}. An RF pulse excites the resonator coupled to the qubit, and the reflected signal encodes the qubit state in its phase. The control hardware digitizes this signal with high-speed ADCs and processes it through a low-latency pipeline to extract a binary outcome~\cite{stefanazzi2022qick,xu2023qubic,liu2025risc}. Because the QEC cycle allows only microseconds per round, all acquisition and processing must finish within that window.

\subsubsection{Syndrome Decoding}

Given the measured syndromes, the decoder infers the most likely physical errors~\cite{wu2022interpretation}. To suppress the effects of measurement noise, each QEC cycle usually consists of multiple rounds of syndrome measurements, decoding therefore operates in both space and time, exploiting correlations across qubits and rounds to separate real data-qubit errors from transient measurement errors. After the errors are obtained, corrections are applied either as physical gate operations or through updates to the logical Pauli frame. Popular decoders include minimum-weight perfect matching, union-find, belief-propagation–based, and neural-network–based methods, each offering different tradeoffs among accuracy, cost, and latency. (Cf. Section~\ref{sec:related}.) 

Real-time feedback from the QEC decoder is crucial~\cite{google2024qec,caune2024demonstrating}. The decoded error information determines the subsequent control operations, particularly for implementing non-Clifford gates which requires mid-circuit branching. Therefore, the throughput and latency of the decoding and feedback pipeline directly affect the fidelity of logical operations.

\textbf{Throughput:} Each QEC cycle generates a continuous stream of syndromes. The decoder must keep pace with this stream and produce outputs without backlog. In advanced superconducting systems like Google Willow, one syndrome round completes in about 1.1 \textmu s~\cite{google2024qec}, leaving little slack for aggregation, decoding, and feedback.

\textbf{Latency:} The decoded result must be available before the next quantum instruction that depends on it. High latency not only delays feedback but also allows errors to accumulate across cycles. 
Achieving both high throughput and low latency is challenging because decoding is computationally intensive and hardware resources are limited. While high-speed RF acquisition~\cite{stefanazzi2022qick,xu2023qubic,liu2025risc} and scalable decoders~\cite{liyanage2024fpga,wu2023fusion,wu2025micro} have been demonstrated individually, integrating them into a complete, end-to-end real-time QEC control stack remains an open engineering challenge.

For completeness, we note that full-cycle QEC workloads involve more than syndrome extraction and decoding~\cite{RevModPhys.87.307,preskill1997faulttolerantquantumcomputation}. Maintaining a logical qubit requires continuous Pauli-frame management and support for fault-tolerant logical operations such as lattice surgery or magic-state distillation. These functionalities place additional constraints on the control architecture, especially on timing predictability, memory organization, and classical computation. While the present system focuses on real-time syndrome decoding and low-latency feedback, extending the architecture to fully support logical operations is an important direction for future work.

\section{System Design}

\begin{figure*}[ht]
  \centering
  \includegraphics[width=\linewidth]{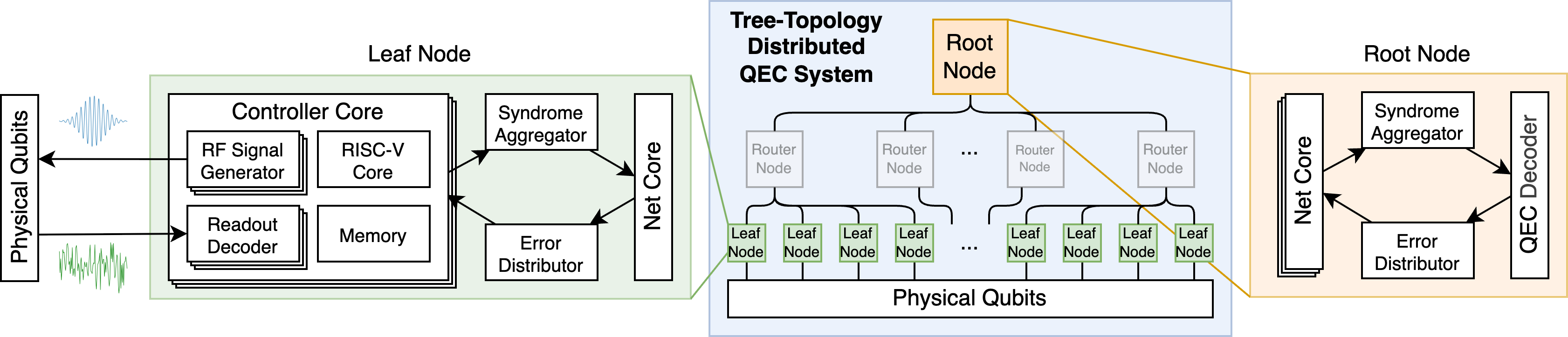}
  \caption{The tree topology of the distributed architecture. The leaf nodes are connected to the physical qubits and process RF signals for controlling qubits. The root node hosts the QEC decoder that decodes syndromes in real-time. The system is extensible by introducing more router nodes.}
  \label{fig-qec-arch-overview}
\end{figure*}

While existing quantum control solutions~\cite{stefanazzi2022qick,xu2023qubic} satisfy the stringent real-time requirements of small superconducting systems---typically managed by one or, at most, two FPGA boards---scalable quantum error correction requires a control architecture capable of executing a vast number of time-critical operations across many qubits, far beyond the capacity of conventional single-core or single-board controllers. This necessitates a scalable multi-core, multi-board architecture with optimized communication and synchronization to meet stringent real-time requirements.

Specifically, as shown in Fig.~\ref{fig-qec-arch-overview}, our design adopts a hierarchical architecture consisting of a multi-core controller, where each qubit is managed by a dedicated RISC-V–based control core, and a multi-board network, which interconnects multiple controllers through a low-latency fiber network. We demonstrate that our architecture provides the desired parallelism, scalability, and synchronization for real-time QEC across hundreds or thousands of qubits.

\subsection{Multi-core Controller}

Practical quantum error correction protocols demand extremely high physical operation throughput, which motivates a multi-core control architecture. 
For instance, in recent superconducting qubit experiments, single- and two-qubit gate durations range from 18ns to 42ns \cite{google2024qec}. To keep pace, the controller must be capable of issuing a new gate roughly every 18ns per qubit---equivalent to 55.6 million operations per second. Existing open-source control systems\cite{stefanazzi2022qick,xu2023qubic} operate near 500~MHz, meaning that for each qubit, the next gate must be prepared and launched within roughly 9 clock cycles. As systems scale toward thousands of qubits, this implies that hundreds of control actions must be issued every clock cycle, quickly exceeding the instruction throughput of any single-core controller.

To meet these throughput and scalability requirements, we adopt a \emph{multi-core} architecture in which each qubit is assigned a dedicated RISC-V–based controller core. A similar approach was used in QubiC~\cite{xu2023qubic}, where each qubit was paired with a dedicated core featuring 128-bit custom instructions. Although effective, QubiC's design relied on a custom ISA that complicates programming and software tooling. In contrast, our design leverages standard RISC-V cores, benefiting from a mature compiler toolchain, debugging ecosystem, and extensibility, all without compromising real-time performance.

We also adopt a \emph{distributed} memory architecture because memory organization is a primary factor governing latency determinism. The alternative shared memory introduces contention and unpredictable access delays albeit providing a more convenient programming model. In conventional processors, multi-level caches mask this variability; however, cache misses routinely introduce penalties above 100 ns, far too large for superconducting qubit control without inducing jitter. Thus, consistent with the design philosophy of QubiC~\cite{xu2023qubic}, each control core in our system is equipped with its own local memory, guaranteeing deterministic access and eliminating contention. Although this reduces programming flexibility compared to a shared-memory design, it is required to sustain precise timing and the throughput demands of real-time QEC.

Together, these design choices yield a flexible, deterministic, and scalable control architecture capable of supporting large-scale QEC experiments without sacrificing synchronization or instruction throughput.

\begin{figure}[ht]
  \centering
  \includegraphics[width=\columnwidth]{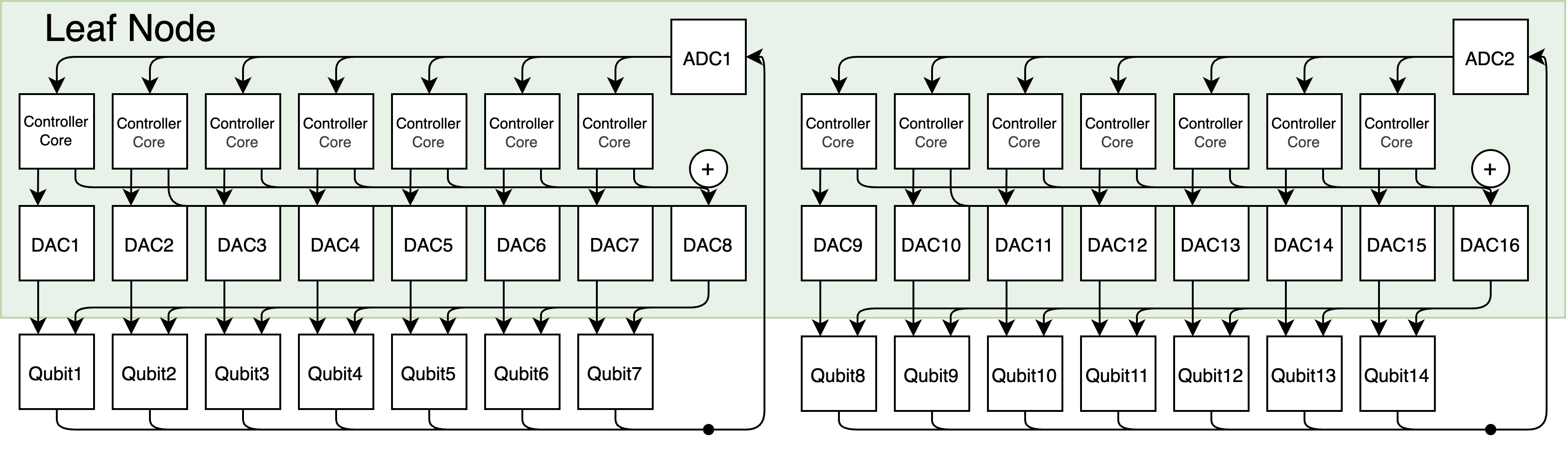}
  \caption{A typical setup of a 14-qubit control system on a AMD ZCU216 RFSoC. Each qubit has a dedicated RISC-V controller, associated with a DAC channel (DAC1-7 and DAC9-15) for gate operation. Besides, the qubits are grouped in 7, where all qubits in a group share a DAC channel (DAC8 and DAC16) for measurement.}
  \label{fig-setup-zcu216}
\end{figure}

\subsection{Multi-board Controller}

Although a single controller board can host multiple control cores and RF channels, the number of high-speed DACs is fundamentally constrained by hardware resources, making a multi-board architecture necessary for scaling to hundreds or thousands of qubits.

Each superconducting qubit typically requires at least two RF outputs: (1) a dedicated gate channel for generating precisely calibrated control pulses, and (2) a readout drive channel, often shared within a qubit group. Flux-tunable qubits may require an additional flux-bias channel~\cite{silva2025manarat}.

However, no single controller board can provide thousands of RF outputs. For example, the AMD ZCU216 RFSoC---widely used in modern quantum control systems~\cite{xu2023qubic,liu2025risc}---offers only 16 high-speed DACs. After reserving two channels for measurement drives, at most 14 qubits can be controlled per board (Fig.~\ref{fig-setup-zcu216}). This limitation is not unique to FPGAs: the number of DAC channels on any integrated platform is fundamentally constrained by power consumption, available bandwidth, routing density, and thermal budgets~\cite{malinowski2023wire}. As these physical constraints persist even in custom ASIC designs, scaling to hundreds or thousands of qubits necessarily requires a multi-board architecture that aggregates RF channels across many controller nodes.

To maintain real-time behavior in multi-board systems, two capabilities are essential: \emph{low-latency communication} and \emph{high-precision synchronization}.

We implement low-latency communication using a custom fiber-optic network that distributes control and measurement data across boards. As illustrated in Fig.\ref{fig-qec-arch-overview}, the system adopts a hierarchical tree topology. Leaf nodes are controller boards connected directly to qubits; the root node hosts the QEC decoder; and intermediate router nodes aggregate and forward data. This design offers modular scalability---new routers and leaves can be added without redesigning lower-level logic---and minimizes communication distance, which is crucial for maintaining sub-microsecond feedback latency\cite{zhao2025distributed}.

High-precision synchronization is achieved by equipping each board with a globally synchronized timer. Synchronization is maintained through a lightweight implementation of Precision Time Protocol (PTP)~\cite{eidson2002ieee}, ensuring sub-nanosecond alignment across all nodes. This guarantees coherent multi-qubit pulse execution and preserves phase relationships necessary for entangling gates.

Together, these principles extend the controller architecture from a single board to a distributed cluster of tightly synchronized control modules, enabling scalable real-time operation across thousands of qubits while meeting the strict latency and coherence requirements of fault-tolerant quantum processors.

\subsection{QEC Decoding Process}

The QEC decoder is centralized at the root node to minimize communication latency and simplify system-wide coordination.
Placing the decoder at the root of the controller network ensures that stabilizer syndrome data from all controller boards reach the decoding engine through the shortest possible paths. This centralized strategy reduces overall feedback delay and provides a clean, architectural interface: the decoder functions as an independent module that can be swapped with alternative implementations, such as minimum-weight perfect matching~\cite{wu2023fusion,wu2025micro}, union-find~\cite{liyanage2024fpga}, or belief-propagation-based decoders~\cite{maan2025decoding,maurer2025real}, without modifying the rest of the control system.

Syndrome data are compacted and transmitted upward through a hierarchical network for efficient aggregation.
At the end of each measurement round, every leaf node (i.e., the controller board directly connected to the qubits) compacts its stabilizer outcomes into a syndrome message. These messages contain the binary parity results extracted from local ancilla qubits. The fiber-optic network then forwards these messages through intermediate router nodes, which aggregate data from multiple child nodes. This hierarchical design supports high-throughput, parallel syndrome collection across many boards.

Decoding occurs at the root node, and the resulting corrections are distributed back down the hierarchy.
Once the root node has received all syndrome messages, it launches the decoding procedure to infer the most likely physical errors on the data qubits. After computation, the root node generates error messages targeted to specific leaf nodes. These results are then propagated downward through the same multi-level network. Upon arrival, each leaf node’s local controllers use the decoded errors to determine the next control actions, including mid-circuit branching and updates to the Pauli frame. The entire feedback loop is illustrated in Fig.~\ref{fig-latency}.

For larger systems that exceed the capacity of a centralized hardware decoder, decoding can be offloaded to GPUs \cite{caldwell2025platform} or implemented using distributed architectures \cite{liyanage2025network}, at the cost of additional communication overhead. 

These mechanisms together integrate the controller, network, and decoder into a scalable real-time QEC architecture.
Three tightly coupled layers enable this integration: (1) a multi-core controller providing deterministic real-time per-qubit control; (2) a multi-board, fiber-linked network supporting low-latency communication and global synchronization across channels; and (3) a centralized, hardware-accelerated decoder performing real-time QEC feedback. Collectively, these components create a coherent and latency-optimized architecture capable of supporting scalable quantum error correction on large superconducting processors.

\section{Implementation}
Implementing a prototype of our real-time distributed QEC control system must resolve the following critical challenges: 

\textbf{Real-time RF signal processing:} Each controller must generate and process RF signals with cycle-accurate timing under tight real-time constraints to ensure correct gate execution.

\textbf{Low-latency communication:} As syndrome measurements are distributed across multiple boards, the system must rapidly aggregate syndrome data for centralized decoding and return correction results to the corresponding controllers.

\textbf{Multi-board synchronization:} The timers on all controller boards must remain precisely synchronized to maintain phase coherence across gigahertz-frequency control signals.

\textbf{High-throughput and low-latency decoding:} The decoder must finish within one QEC cycle to avoid cascading delays and reduced throughput. 

\textbf{Distributed software infrastructure:} A scalable software framework is required to coordinate distributed control logic, manage calibration and scheduling workflows, and synchronize operations across all boards.

In this section, we describe how we address these challenges via hardware–software co-design and precise synchronization. Our system is implemented using RISC-Q, an open-source hardware generator developed in SpinalHDL that produces customizable RISC-V–based quantum control systems \cite{liu2025risc}. RISC-Q provides a parametric framework for generating the whole controller SoC, including controller cores, memories, interconnect fabrics, and MMIO peripheral interfaces, tailored for real-time quantum control. 
Thanks to RISC-Q’s substantial boost in development efficiency, we are able to develop our whole system with 
14,648 lines of HDL codes, even including a rewrite of the hardware decoder Helios~\cite{liyanage2024fpga} in SpinalHDL, as well as software implementation with 1,521 lines of codes in Python and C. 
The entire process of design, testing, and deployment to real hardware requires about eight person-months.
Notably, this generator-based development approach enables a modular and flexible system implementation, allowing the architecture to scale without a big increase in codebase size.

\subsection{RF Signal Generation with Accurate Timing}

\begin{figure}[ht]
  \centering
  \includegraphics[width=0.7\columnwidth]{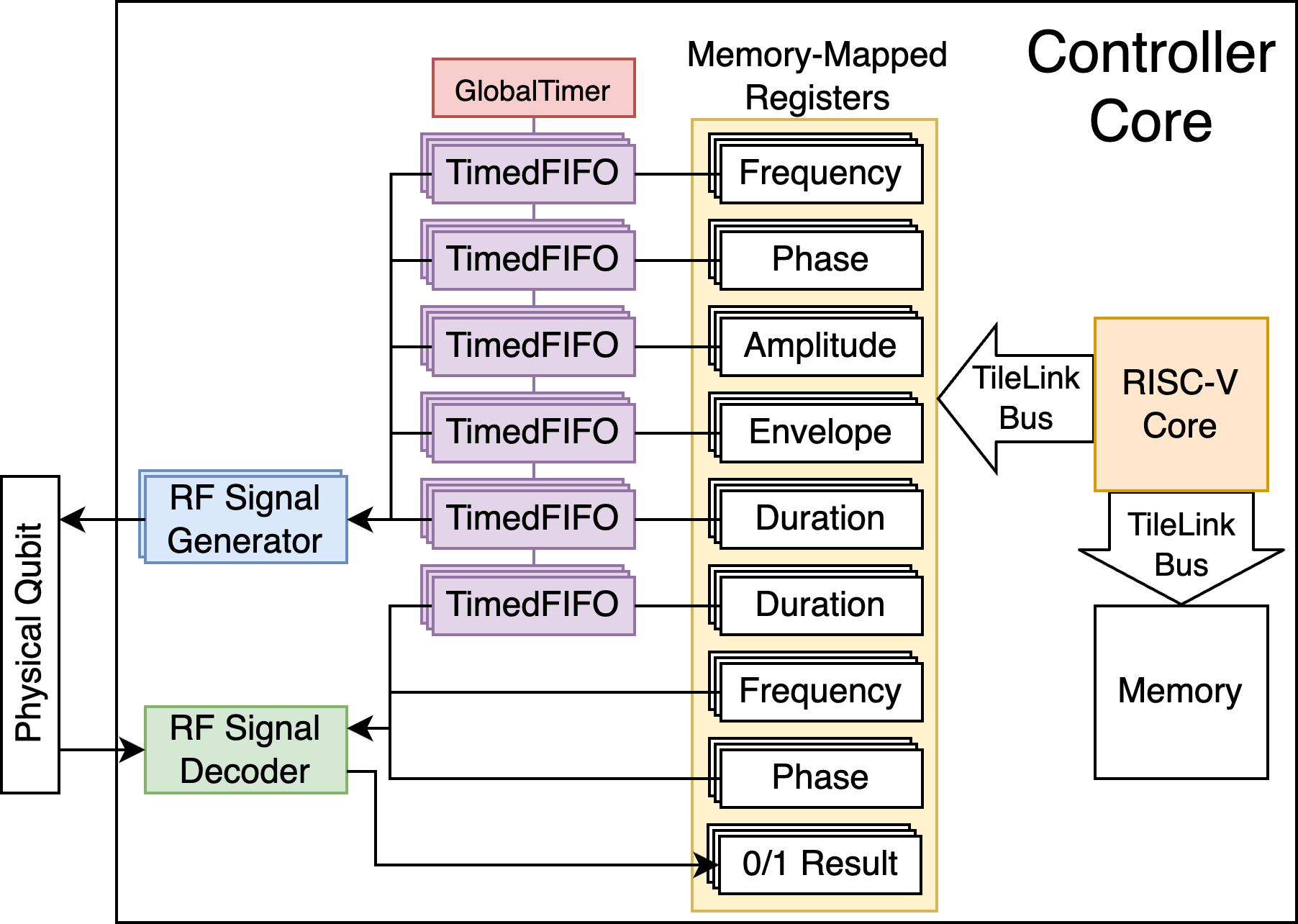}
  \caption{Controller Core: A Schematic Overview}
  \label{fig-qubit-controller}
\end{figure}

Our implementation integrates a DDS–based pulse generator with cycle-level timing control. The main components of a controller core are shown in Fig.~\ref{fig-qubit-controller}. When synthesized on the AMD ZCU216 RFSoC, the controller meets a 500 MHz clock rate, matching the state-of-the-art among open-source quantum control platforms, and provides sufficient timing resolution to support nanosecond-scale pulse scheduling for superconducting qubits.

\subsubsection{DDS-based RF Signal Generator}
Each controller core controls two DDS-based RF signal generators that generate pulses for qubit gates and readout operations. The generator supports configurable phase, frequency, envelope and amplitude, while achieves a sampling rate of 8 gigasamples per second. Every pulse parameter is stored in memory-mapped configuration registers that can be updated at runtime through a TileLink bus. This memory-mapped interface enables the RISC-V cores to directly control pulse parameters using standard load/store instructions, providing both flexibility and low-latency access.

\subsubsection{Timed FIFO and Synchronized Control}
To achieve cycle-accurate pulse generation, each RF signal generator is synchronized to a global timer shared across all controller cores. The connection between the processors and the signal generators is mediated by hardware Timed FIFOs \cite{stefanazzi2022qick}, which buffer pulse parameters produced by the RISC-V cores and release them at precise timestamps determined by this global timer. 
This design decouples processor instruction timing from deterministic pulse timing, ensuring phase-coherent, cycle-accurate alignment across all channels.

\subsection{Network}

To meet the stringent latency and synchronization requirements, we implement a network core directly on the Gigabit Transceivers (GTs) of the RFSoC and employs a custom lightweight protocol based on 64B/66B line encoding \cite{walker200064b}.

\subsubsection{64B/66B-based Physical Layer}

The network leverages the high-speed GT transceivers available on the RFSoC, operating with 64B/66B encoding. Compared with conventional network IPs such as AMD Aurora~\cite{amdaurora}, which introduce additional protocol framing and handshake cycles, our custom implementation based on the Eos core \cite{liyanage2025network,eoscore} adopts a simplified link layer with smaller data units and more frequent error correction codes, allowing data processing in fewer cycles. The reliability of this custom implementation is validated through a 12-hour stress test. Our design achieves a substantial reduction in one-way link latency, from approximately 450 ns with Aurora to about 156 ns.  This improvement directly enhances the responsiveness of the real-time feedback loop in QEC cycles.

\subsubsection{Tree-topology and Scalability}

The inter-board network follows a tree topology, where leaf nodes represent controller boards connected directly to physical qubits, and intermediate routers aggregate data from lower layers and forward it toward the root node, which hosts the centralized decoder. This topology supports modular scalability, allowing new routers to be added to accommodate larger qubit number without altering the low-level board design. However, adding routing layers introduces additional propagation latency proportional to the tree depth. In our prototype implementation, the leaf nodes are connected directly to the root node, achieving minimal communication latency while supporting up to $4 \times 14 = 56$ qubits with 4 AMD ZCU216 RFSoCs serving as leaf nodes and another ZCU216 as the root node.

\subsection{Clock Synchronization via PTP}

Accurate synchronization across boards is essential for maintaining phase coherence in distributed control signals. In our prototype, all controller boards share a common reference clock, minimizing inter-board phase differences. Synchronization is then reduced to aligning the global timer registers on each board, which we achieve using a minimal Precision Time Protocol implementation~\cite{xu2025multi} over the GT network. This provides sub-nanosecond alignment of global timers, enabling coherent execution of cross-board gates and measurement sequences.
For larger-scale systems where a single reference clock is impractical, the architecture can be extended using the White Rabbit protocol~\cite{moreira2009white}, which offers picosecond-level clock and time synchronization over fiber links. White Rabbit allows each board to maintain phase-locked clocks and synchronized global timers over long distances, preserving timing coherence in large distributed quantum control networks.

\subsection{QEC Decoder}

Efficient QEC requires a decoder capable of processing syndrome data and producing correction results with minimal delay, as this latency directly determines the logical clock rate of the system~\cite{caune2024demonstrating}. As a result, achieving predictable low latency is a primary design requirement for the decoder.

\subsubsection{Choosing Hardware Decoders}
While our QEC system can, in principle, support any decoder, we focus on open-source projects to keep the entire system fully open-source. Among these, Helios~\cite{liyanage2024fpga} and Micro Blossom~\cite{wu2025micro} are currently the two highest-performance hardware QEC decoders for the surface code. Both offer scalable designs but differ fundamentally in their hardware–software partitioning and latency characteristics, which affect their suitability for integration.

Micro Blossom adopts a hybrid architecture in which part of the decoding logic runs on an FPGA while the remaining computations are performed on an external CPU. This design provides algorithmic flexibility and facilitates the integration of more complex and higher-accuracy decoding strategies. However, it introduces communication overhead between the FPGA and CPU, making the overall latency dependent on the performance of both devices and their interconnect. Consequently, its timing behavior is less predictable, which limits its suitability for hard real-time feedback loops in QEC systems.

In contrast, Helios is a fully hardware-implemented decoder, optimized for lower and more deterministic latency. The entire union-find-based decoding algorithm is executed directly on the FPGA fabric. By avoiding external CPU interaction, Helios achieves more predictable timing performance, typically on the order of a few hundred nanoseconds\footnote{For example, for decoding the distance-13 surface code, Micro Blossom requires 0.8 \textmu s in average, while Helios requires 250 ns.}, making it well suited for integration into our real-time QEC control pipeline.

\subsubsection{Integration with the Control System}
In our distributed architecture, the Helios decoder runs at the root node of the controller network, receiving syndrome messages from all leaf nodes. A syndrome aggregator assembles these messages into a complete input frame for the decoder. Once all syndromes for a QEC round are collected, Helios begins decoding. When a valid result is produced, an error distributor generates correction messages with the identified error bits for each leaf node and transmits them back through the network interface.
This fully hardware-implemented pipeline ensures low end-to-end latency across the distributed system.

\subsubsection{Scalability and Future Extensions}
While Helios is used in our prototype implementation, the system architecture is designed to be decoder-agnostic. A clear interface defined by the syndrome aggregator and error distributor separates the decoding logic from the rest of the control infrastructure. This modular boundary allows the decoder to be replaced or upgraded without altering the rest part of the controller system.

Alternative hardware decoders, such as those based on belief-propagation~\cite{maan2025decoding,maurer2025real} or neural-network~\cite{overwater2022neural,ataides2025neural}, can thus be integrated with minimal modification. This flexible and extensible interface accommodates rapid advances in QEC decoding research while preserving the real-time performance and determinism required for fault-tolerant operation.

\subsection{Software Implementation}

\begin{figure}[ht]
  \centering
  \includegraphics[width=0.8\columnwidth]{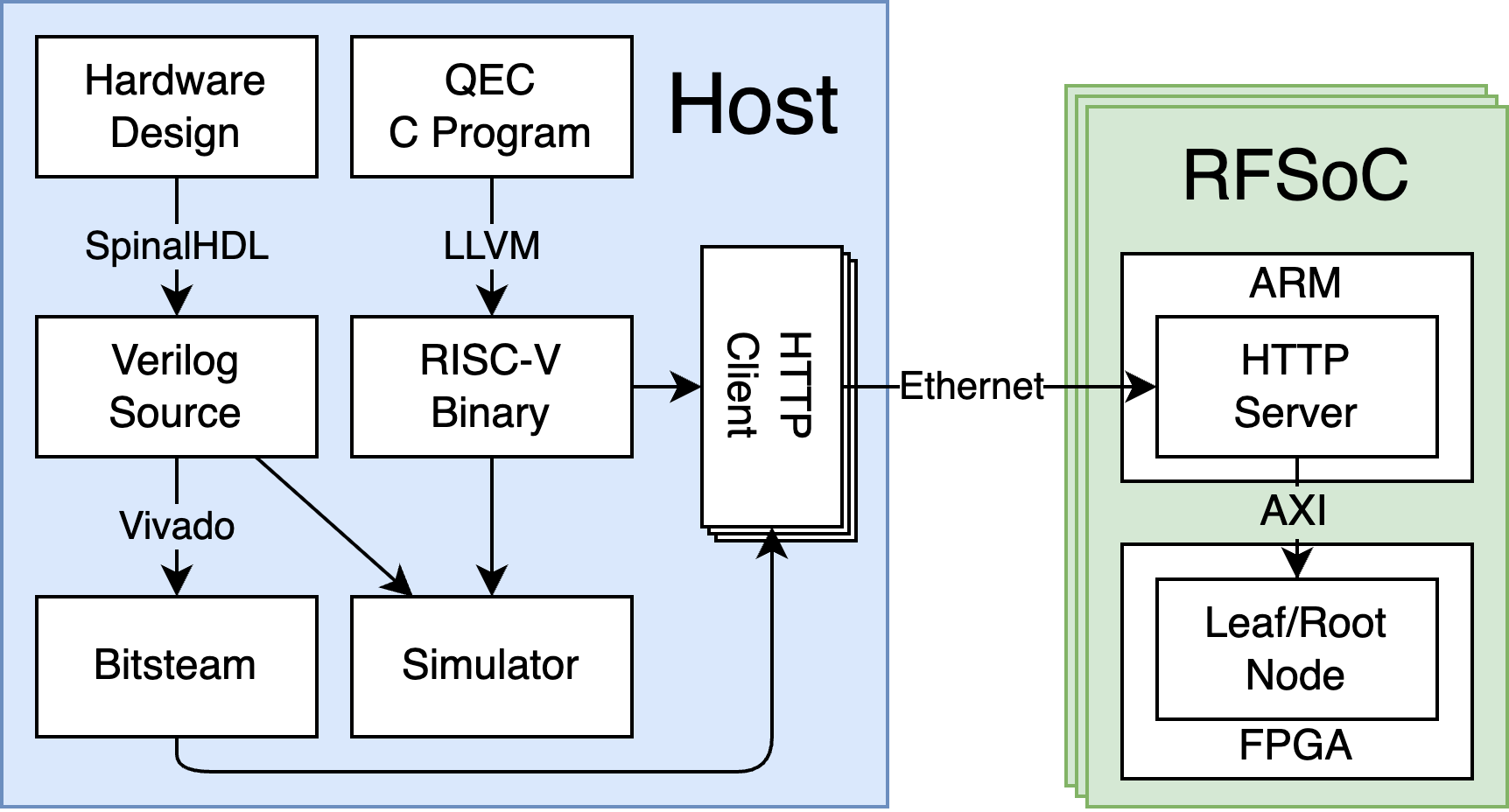}
  \caption{Software stack of the QEC system. Hardware modules are generated in SpinalHDL and synthesized into a bitstream, while control programs are compiled into RISC-V binaries using the standard RISC-V LLVM toolchain. Both artifacts are delivered from the host to the RFSoC over Ethernet via an HTTP interface, where an HTTP server on ARM loads them onto the FPGA fabric.}
  \label{fig-software-stack}
\end{figure}

We implement a software stack (shown in Fig.~\ref{fig-software-stack}) which provides the interface between the user, the distributed controller hardware, and the underlying FPGA fabric, facilitating both system configuration and experimental orchestration.

\subsubsection{Host–Board Communication Framework}

Each controller board hosts an HTTP-based server that exposes configuration and monitoring interfaces for the FPGA fabric. The server runs on the ARM processing system of the RFSoC and is built upon the PYNQ framework, which provides Python bindings for memory-mapped I/O (MMIO) access to the FPGA through the AXI bus. Through this interface, users can configure pulse envelopes, load firmware, and monitor runtime states such as timer offsets and network link status.

We develop a dedicated Python client library to coordinate multiple boards from a host desktop over Ethernet, which provides high-level APIs for remote configuration, bitstream deployment, and trigger synchronization. This design allows the host to treat the entire multi-board system as a unified control fabric while retaining the ability to configure and monitor each board independently. Its HTTP-based interface is language-agnostic and user-friendly, simplifying the development of remote control and automation tools.

\subsubsection{On-Board Real-Time Software}

On each controller FPGA, the real-time control software runs on RISC-V cores, which is readily supported by RISC-Q,  implemented within the FPGA fabric. These programs are written in C and compiled using the standard RISC-V LLVM toolchain, leveraging the extensive software ecosystem built around the open RISC-V ISA. This enables developers to use mature compilers and debugging tools without relying on custom toolchains or proprietary workflows. By building the control architecture on RISC-Q, we gain both portability across hardware configuration and a stable, well-supported programming interface, making application development and long-term maintainability significantly simpler than in custom-ISA control systems. 

The software implements the quantum circuit execution for QEC by coordinating the complete RF signal processing workflow: generating pulses for entangling gates and syndrome measurements, decoding readout signals to obtain syndromes, writing the results to the syndrome aggregator, and retrieving error data from the error distributor. All interactions between the RISC-V cores and peripherals, including the RF signal generators, decoder, syndrome aggregator, and error distributor, are implemented through MMIO, providing a straightforward interface.

The same software binary can also be executed on a cycle-accurate hardware simulator that we build in SpinalHDL and compile with Verilator. This simulator includes all core components of the QEC system, including the RISC-V cores, MMIO interfaces, RF signal processors, and the Helios decoder, allowing developers to validate system behavior, timing correctness, and integration logic before deploying to the physical RFSoC boards.

\subsubsection{System Integration and Extensibility}

Together, the host–board communication framework and the on-board real-time software form a hierarchical software stack spanning user-level experiment orchestration to hardware-level control. This modular structure simplifies system scaling, where new boards can be added and configured through the same software interface without modifying the low-level firmware.

By separating configuration logic (Python/PYNQ layer) from real-time execution (C/RISC-V layer), the software architecture combines flexibility for researchers with deterministic performance for hardware, enabling a practical and scalable development environment for distributed quantum control.

\section{Evaluation}
\label{sec-evaluation}

We evaluate the performance of the proposed real-time QEC control architecture through FPGA implementation and multi-board experiments. The evaluation focuses on \emph{three} key aspects: hardware resource utilization and achievable frequency, end-to-end latency and the throughput of the QEC process.

\subsection{System Implementation and Measures}

\begin{figure}[ht]\centering
\includegraphics[width=0.8\columnwidth]{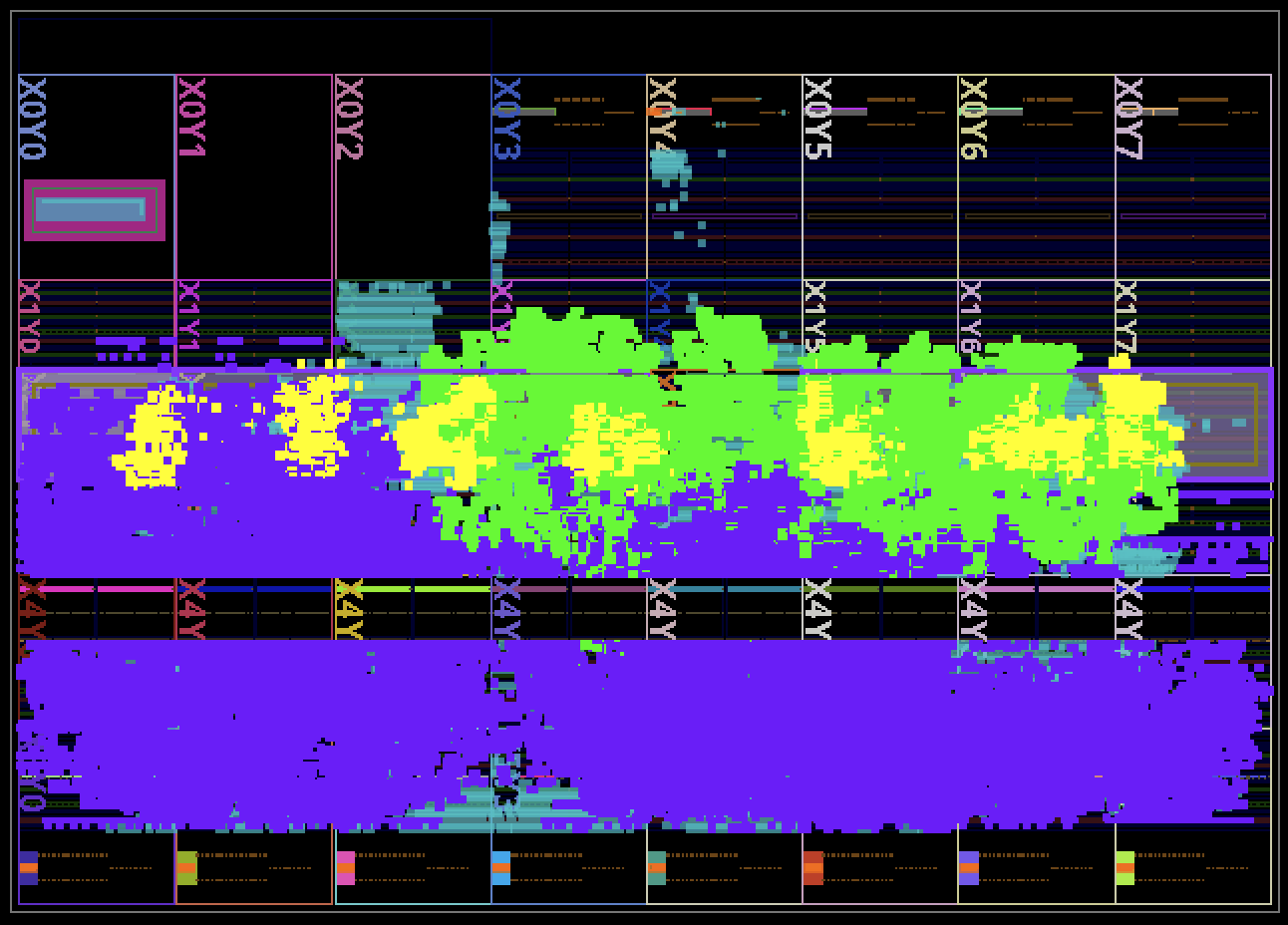}
\caption{FPGA floorplan of a leaf-node implementation on the AMD ZCU216 RFSoC. 
Yellow areas correspond to RISC-V controller cores, green areas to RF signal decoders, and purple areas to RF signal generators.}
\label{fig-fpga-floorplan}
\end{figure}

Our implementation is synthesized targeting the AMD ZCU216 RFSoC as the controller board in the distributed architecture. Each leaf node in the network is configured to control up to 14 qubits, as demonstrated in Fig.~\ref{fig-setup-zcu216}. The FPGA resource utilization and maximum operating frequency of the design are summarized in Table \ref{tab-utils-freq}, with the floorplan illustrated in Fig.~\ref{fig-fpga-floorplan}. The reported results include utilization of LUTs, flip-flops, block RAMs (BRAMs), and DSP slices for the main components such as RISC-V cores, RF signal generators, RF signal decoders and the network interface. The RF signal generators and decoders dominate the overall footprint, primarily due to the DDS logic. In contrast, the 14 RISC-V controller cores account for a relatively small fraction of LUTs and FFs, indicating that the multicore control architecture scales efficiently with qubit count. 

\begin{table}[ht]
  \scriptsize
  \centering
  \caption{FPGA Resource Utilization and Operating Frequency of Leaf Node Components on AMD ZCU216 RFSoC.}
  \label{tab-utils-freq}
  \begin{tabular}{lccccc}
    \toprule
    \textbf{Module} & \textbf{LUTs} & \textbf{FFs} & \textbf{BRAMs} & \textbf{DSPs} & \textbf{Fmax (MHz)} \\
    \midrule
    ($14 \times$) RISC-V cores & 11228 & 11508 & 28 & -- & 500 \\
    ($28 \times$) RF generators & 63644 & 156604 & 448 & 1372 & 500 \\
    ($14 \times$) RF decoders & 28378 & 95592 & -- & 700 & 500 \\
    Network core & 224 & 450 & -- & -- & 156.25 \\
    \midrule
    \textbf{Total (Leaf Node)}  & 127249 & 266372 & 476 & 1722 & -- \\
    \bottomrule
  \end{tabular}
\end{table}

The design meets timing closure at 500~MHz, matching the typical control frequency required for superconducting qubit systems \cite{xu2023qubic,liu2025risc}. 

\begin{figure}[ht]
\centering
\begin{subfigure}[b]{0.6\columnwidth}
\centering
\includegraphics[width=\linewidth]{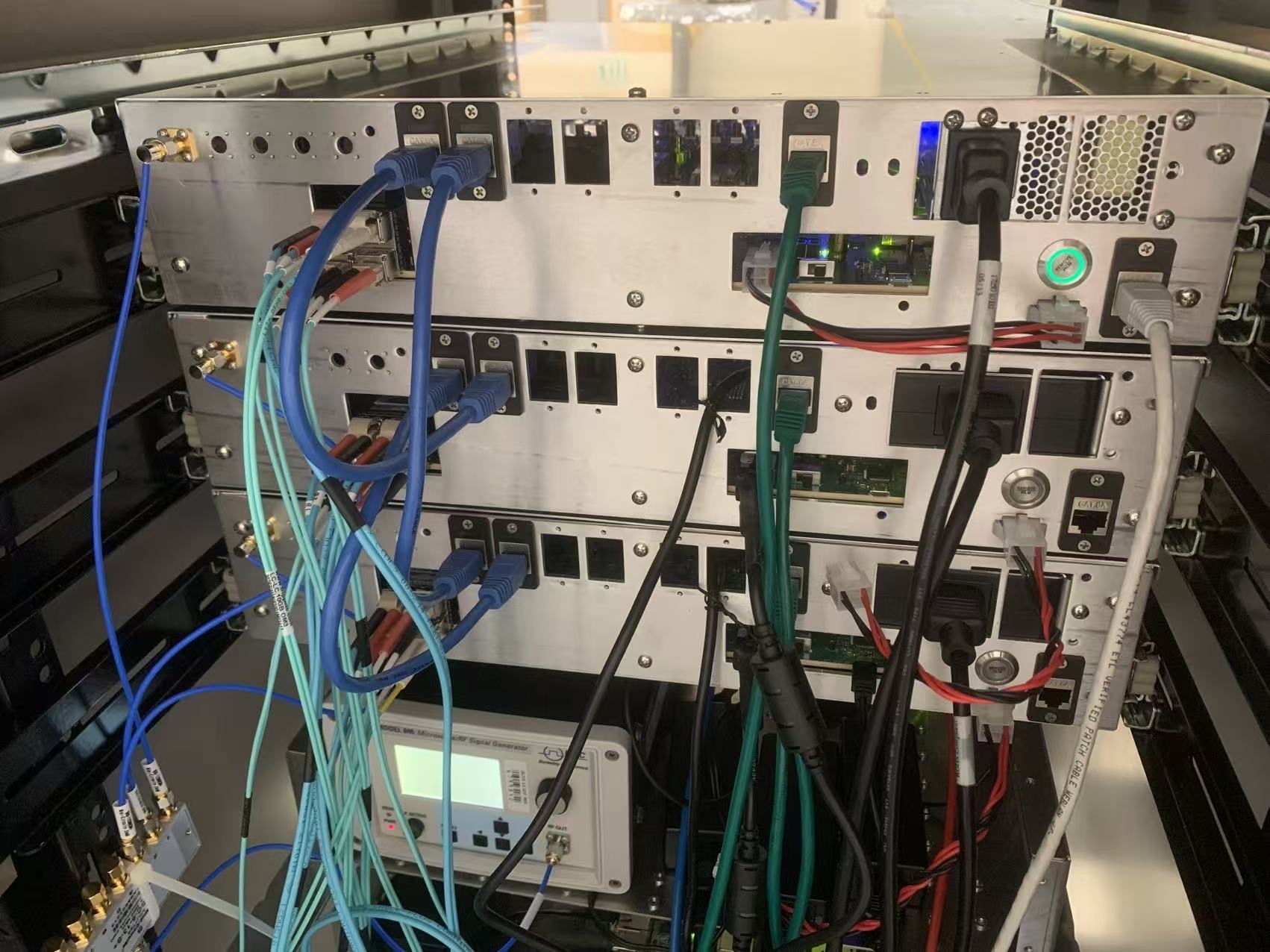}
\subcaption{}
\end{subfigure}
\begin{subfigure}[b]{0.34\columnwidth}
\centering
\includegraphics[width=\linewidth]{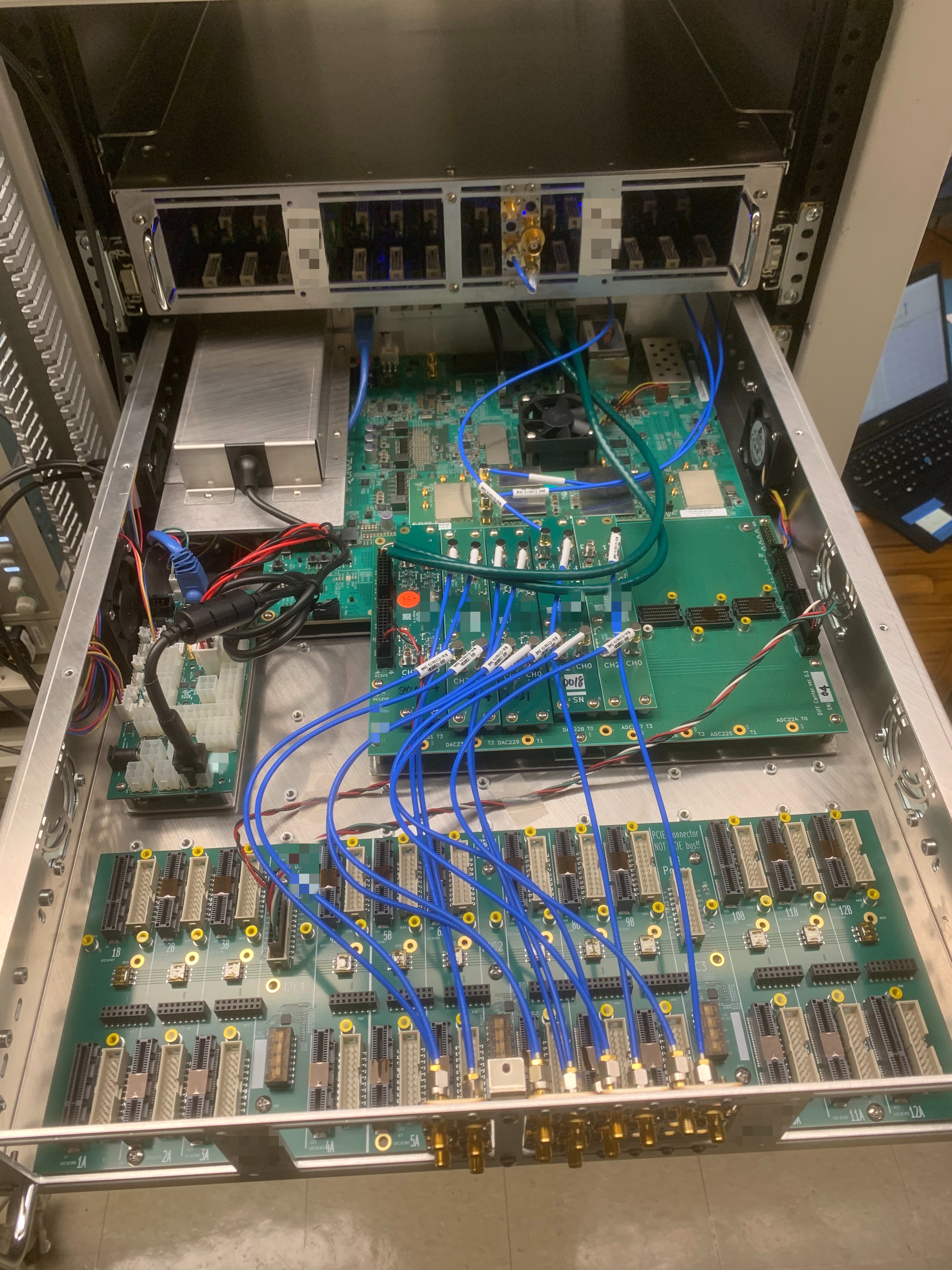}
\subcaption{}
\end{subfigure}
\caption{(a) Three-board experimental setup using AMD ZCU216 RFSoC evaluation boards hosted in separate QubiC chassis. (b) Components inside a single chassis: the ZCU216 board, analog front-end board, and power management board.}
\label{fig-3-board-setup}
\end{figure}

To verify the hardware readiness for real-qubit operation, we evaluate the gateware implementation on three AMD ZCU216 RFSoC boards deployed within the open-source QubiC hardware system~\cite{xu2023qubic}. The setup is shown in Fig.~\ref{fig-3-board-setup}. The QubiC system provides the necessary infrastructure for laboratory deployment, including analog front-end boards for signal amplification and regulation, a power management board for stable power distribution, and a custom chassis that houses and interconnects all RFSoC boards.

The measured root-mean-square (RMS) jitter of the system clock is 1.28~ps (integrated from 100Hz to 100MHz), which fully satisfies the stability requirements for superconducting qubit experiments. The effectiveness of this hardware platform has also been validated in prior experimental work~\cite{xu2023qubic}.

\subsection{Real-Time Performance Evaluation}

To assess real-time performance, the experimental setup shown in Fig.~\ref{fig-3-board-setup} is used, where one board functions as the root node hosting the QEC decoder, and two boards act as leaf nodes connected via the fiber network. This configuration corresponds to the minimal setup required for a distance-3 surface code, enabling realistic timing measurements for the end-to-end QEC control cycle. (Cf., Fig.~\ref{fig-qec-arch-overview} for general system)

\begin{figure}[ht]
\centering
\includegraphics[width=\columnwidth]{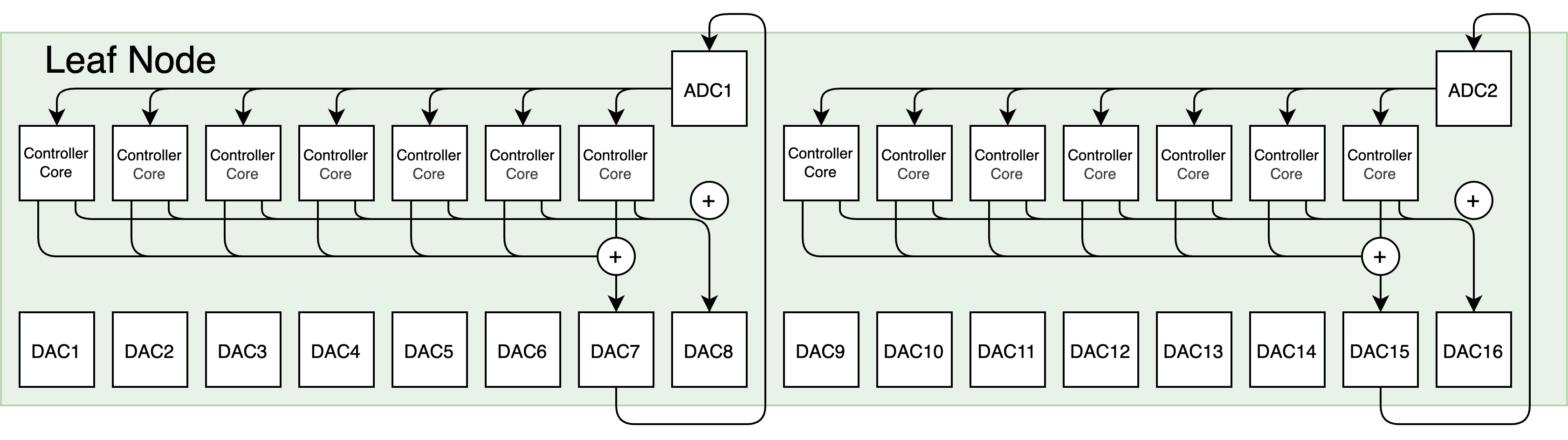}
\caption{Evaluation setup without qubits. The outputs of RF signal generators for gate operations are summed and looped back to the ADC inputs to emulate qubit readout signals.}
\label{fig-loopback-dac}
\end{figure}

We evaluate the real-time performance of the QEC system without physical qubits by looping the RF gate outputs back into the readout ADCs to emulate qubit measurement signals, as shown in Fig.~\ref{fig-loopback-dac}. This loopback setup preserves the actual dataflow and timing characteristics of the control pipeline, enabling accurate latency characterization.

\subsubsection{Latency}
As there is no universally defined QEC cycle time, we focus on measuring the \emph{decoding-feedback latency}--—the time elapsed from when the final 0/1 syndrome results become available on the leaf nodes to when the decoded error information is returned from the root node and ready for feedback control. Using globally synchronized timers across all boards, latency samples were collected over \emph{10,000} experimental runs, capturing all major pipeline stages including syndrome aggregation, network communication, QEC decoding, and error distribution.

Each experiment emulates three rounds of syndrome measurements, matching a full distance-3 surface-code cycle. For each round, the controller cores generate emulated readout signals using the on-board RF signal generator; these signals are looped back into the readout decoder to produce syndrome bits. To ensure the measurement reflects a worst-case decoding scenario, the generated signals are chosen such that the resulting syndrome corresponds to the most time-consuming decoding pattern for the distance-3 Helios decoder.

Once a emulated measurement completes, the syndrome is processed through the leaf-side hardware pipeline. The syndrome aggregator packages the results with a latency of 29 ± 3 ns, after which the message is transferred from the leaf node to the root node through the fiber network in 157 ± 16 ns. At the root node, the incoming data undergo 20 ± 10 ns of aggregation and pre-decode processing before being delivered to the Helios decoder. Under the chosen worst-case pattern, the decoder itself completes in 56 ns.

The decoded errors are then sent back to the leaf nodes, requiring another 25 ± 3 ns for root-side error distribution, 155 ± 9 ns of network transfer, followed by 9 ± 1 ns for leaf-side error distribution. These ± intervals represent the observed min–max spread across trials and arise primarily from cross-clock-domain synchronization between components.

\begin{figure}[ht]
\centering
\includegraphics[width=\columnwidth]{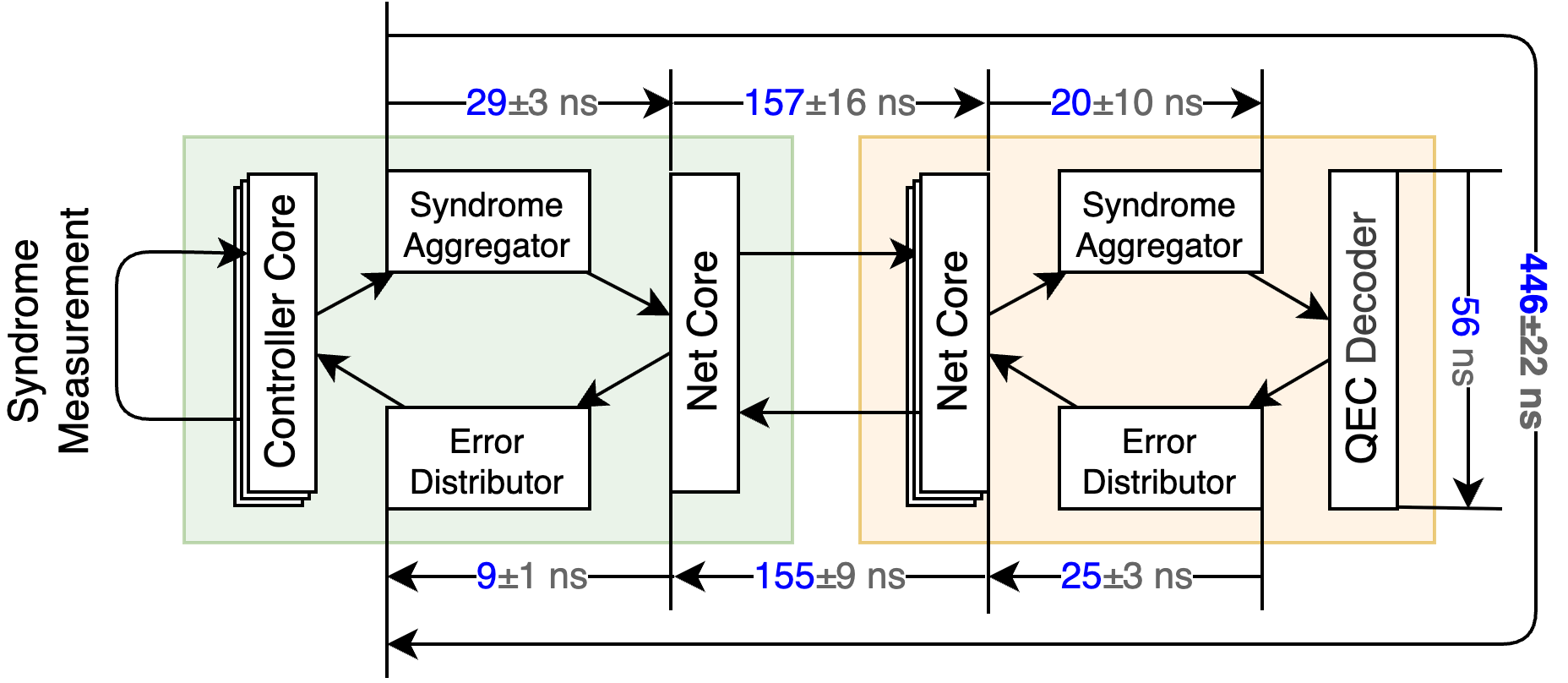}
\caption{Measured average latencies across each stage of the QEC decoding pipeline. The end-to-end decoding-feedback latency for the distance-3 surface code is approximately 446 ns.}
\label{fig-latency}
\end{figure}

Summing all stages yields an average end-to-end decoding-feedback latency of 446 ns, well below the sub-microsecond requirement for superconducting-qubit QEC. All the latencies are summarized in Fig.~\ref{fig-latency}.

Although a full-scale evaluation involving hundreds of qubits would require more hardware than is currently available, the total system latency can be predicted from the measured performance of individual subsystems due to our modular design. 
In the present experiments, all non-decoder components contribute approximately 390~ns of latency.

The average decoding latency of the Helios hardware decoder for surface codes with distances ranging from 3 to 21, implemented on an AMD VCU129 FPGA, has been reported in \cite{liyanage2024fpga}.
These measurements assume a physical error rate of 0.001, consistent with current superconducting qubit performance~\cite{google2024qec}.
Under these conditions, Helios achieves a logical error rate of $10^{-6}$ at distance 11.
Moreover, each AMD VCU129 FPGA board integrates 34 high-speed GT transceivers, allowing it to connect directly to up to 34 leaf nodes, and thus control at most $34 \times 14 = 476$ qubits when used as the root node.

According to \cite{liyanage2025network}, a router node implemented on an AMD VMK180 FPGA can interconnect up to 29 child nodes, enabling large-scale hierarchical expansion.
Based on the latency measured in our evaluation, extending the system with one router layer introduces an additional 45 ns of on-board processing latency for syndrome aggregation and error distribution, and approximately 312 ns of round-trip network latency for inter-node communication.

\begin{figure}[ht]
\centering
\includegraphics[width=\columnwidth]{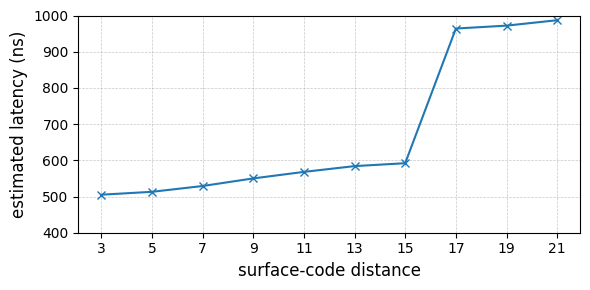}
\caption{Estimated total decoding-feedback latency as a function of surface code distance, extrapolated using reported Helios decoder performance~\cite{liyanage2024fpga} and latency measurements from this work. The configuration assumes AMD VCU129 FPGA as the root node and AMD VMK180 FPGAs as router nodes.}
\label{fig-estimated-latency}
\end{figure}

By combining these empirical and reported results, we estimate the total decoding delay for larger surface code distances, as illustrated in Fig.~\ref{fig-estimated-latency}.
The analysis indicates that the proposed system can scale to distance 21 ($\sim$881 physical qubits) with a single additional router layer, while maintaining an overall end-to-end latency of under 1 \textmu s.
The sharp increase in latency between code distances 15 and 17 arises from the introduction of this router layer: without routing, the system can control up to 476 qubits, whereas a distance-17 surface code requires 577 qubits, exceeding this limit.

\subsubsection{Throughput}
Throughput is a critical parameter for real-time QEC systems, as it determines whether the control pipeline can process all measurement results within each QEC cycle. If the throughput of any components falls below the required syndrome data rate, a backlog will accumulate, delaying subsequent syndrome measurement rounds and ultimately invalidating continuous QEC process.

In the proposed architecture, two components primarily determine throughput performance:
\begin{itemize}
\item \textbf{Network throughput}, which limits how quickly syndrome data can be transferred between leaf nodes and the root decoder.
\item \textbf{Decoder throughput}, which limits how fast syndromes can be processed into error results.
\end{itemize}

In our current prototype, the ZCU216 root node employs four transceivers, each operating at a 10 Gb/s line rate. After accounting for the 64B/66B line encoding overhead (3.03\%), the effective aggregate throughput is $4 \times 10 \times 0.9697 = 38.788$ Gb/s, which corresponds to approximately 38,788 syndrome bits per microsecond that can be transferred from all the leaf nodes to the root nodes.

Although our implementation uses 10 Gb/s transceivers, the ZCU216 board supports transceivers up to 28 Gb/s. With commercially available 28 Gb/s transceivers installed, the same four links would provide up to
$4 \times 28 \times 0.9697 = 108.6$ Gb/s, offering nearly $3 \times$ of network throughput without modification to the control or decoding architecture.

The number of syndrome bits per round depends on the surface code distance. A distance-3 code produces 8 syndrome bits per round, while a distance-21 code produces 440 bits. According to \cite{liyanage2024fpga}, the Helios decoder achieves increasing throughput with code distance, peaking at $d=21$, where 440 syndrome bits are processed in average of 11.5~ns, yielding a raw decoding throughput of $440 / 11.5 = 38.26$ Gb/s.

Combining the network and decoder results, the overall system can theoretically achieve a syndrome processing throughput of 38.26~Gb/s, indicating that the decoder, not the network, is the limiting component.  Considering a typical measurement cycle time of 1 \textmu s, which corresponds to 440 syndrome bits per microsecond (440 Mb/s), the available throughput exceeds the system requirement by nearly two orders of magnitude. This margin ensures that the architecture can sustain continuous, real-time QEC operation up to at least distance-21 surface codes (881 physical qubits) without risk of backlog or timing violation.

For larger code distances, additional routing layers may be introduced to aggregate syndrome data from more leaf nodes, which can increase network latency due to serialization. However, if the AMD VCU129 FPGA---used in the original Helios implementation~\cite{liyanage2023scalable}---is deployed as the root node instead of the ZCU216, it can directly connect to up to 34 router nodes. In this configuration, each router node only needs to process
$440 / 34 \approx 13$
syndrome bits per round. This fits within a single 64-bit data frame of our network core, eliminating the need for serialization and thereby avoiding additional serialization latency.

\subsection{Comparison with Existing Integrated QEC Systems}

To contextualize the performance of our system, we compare its real-time QEC latency with two existing integrated QEC demonstrations: Google’s system~\cite{google2024qec} and the Rigetti--Riverlane (R\&R) system~\cite{caune2024demonstrating}. Although these systems differ in architecture and experimental configuration, they all implement the core tasks required for real-time QEC:
\begin{itemize}
  \item \textbf{Syndrome aggregation}: transferring syndrome data from the qubit-control hardware to the decoder;
  \item \textbf{QEC decoding}: computing the most likely physical errors from the measured syndromes;
  \item \textbf{Error distribution}: sending decoded errors back to the qubit controller for feedback (implemented by R\&R and our system, but not by Google).
\end{itemize}

\begin{figure}[ht]
\centering
\includegraphics[width=\columnwidth]{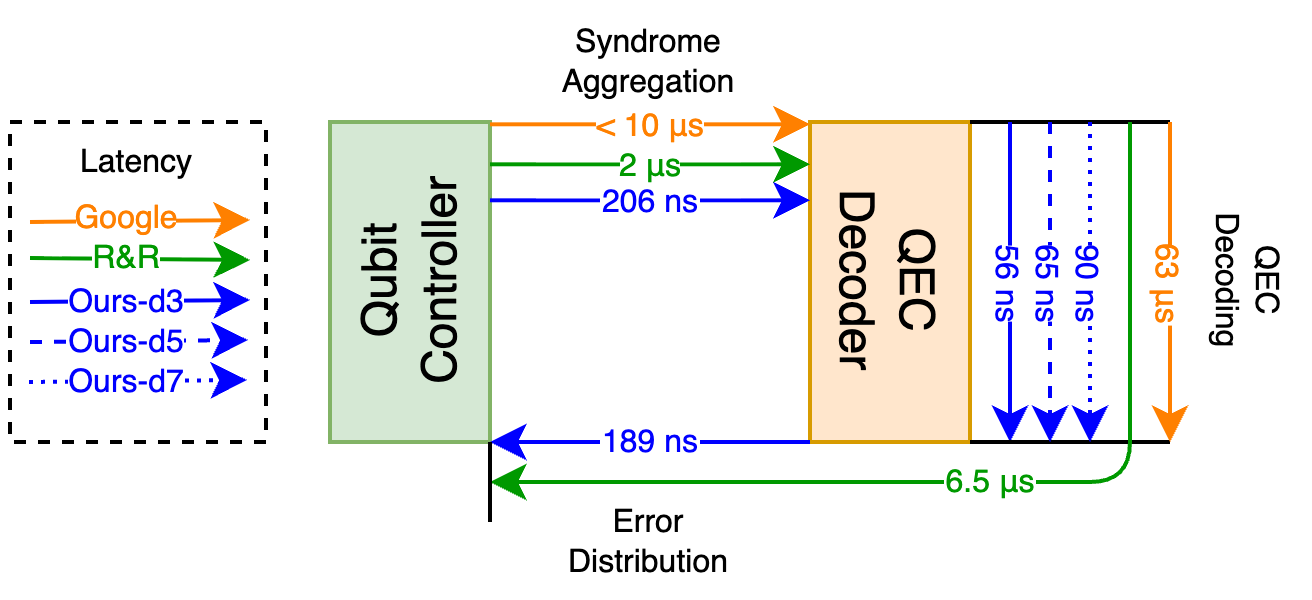}
\caption{Comparison of syndrome aggregation, QEC decoding, and error distribution latency across integrated QEC systems: Google~\cite{google2024qec}, Rigetti--Riverlane (R\&R)~\cite{caune2024demonstrating}, and this work. 
The end-to-end decoding–feedback latency of our system is evaluated for a distance-3 surface code. The standalone decoding latency for distance-5 and distance-7 is measured by feeding pre-sampled syndrome data (with physical error rate $10^{-3}$) directly to the Root Node.
}
\label{fig-latency-comparison}
\end{figure}

Our architecture achieves a total QEC decoding-feedback latency of 446 ns for a distance-3 surface code with 3 syndrome measurement rounds, decomposed as 206 ns of syndrome aggregation, 56 ns of QEC decoding and 189 ns of error distribution. All three stages are implemented as custom FPGA datapaths with no host CPU in the critical loop, which minimizes software and off-board communication overheads. 

We further evaluate decoding performance for larger codes by feeding pre-sampled syndrome data directly to the Root Node. When the physical error rate is 0.001, the average decoding latency is 65~ns for distance-5 and 90~ns for distance-7 surface codes.
End-to-end decoding–feedback latency is not reported for these larger code distances because the current prototype does not include enough FPGA boards to instantiate the full set of leaf nodes required by the distributed architecture. Nevertheless, the standalone decoding measurements capture the primary scaling component of the QEC pipeline, since aggregation and feedback latency remain largely constant while the decoder latency varies with code distance.

Google’s system is dominated by off-board software decoding, reporting a 63 \textmu s decoding latency for a distance-5 surface code with 10 syndrome measurement rounds on 49 qubits \cite{google2024qec}. In their architecture, syndrome data are streamed from the qubit controller to a  workstation running a real-time Linux system, introducing near 10 \textmu s of communication latency. This design enables algorithmic flexibility but introduces substantial latency from network transfer and CPU processing. Moreover, the experiment does not include a closed feedback loop back to the qubit controller, so no error-distribution latency is reported to be directly compared.

R\&R’s system reduces decode time with an FPGA decoder, but incurs extra aggregation and feedback latency. The Rigetti-Riverlane experiment demonstrates an 8-qubit stability protocol with 9 syndrome measurement rounds and reports a 6.5 \textmu s decode-and-feedback response time~\cite{caune2024demonstrating}. Their decoder is implemented on an FPGA~\cite{barber2025real}, but the syndrome aggregation and error distribution paths traverse multiple system boundaries (Rigetti control hardware, Riverlane FPGA decoder, and intermediary interfaces). These additional translation layers contribute to the reported $\sim$2 \textmu s aggregation time and 6.5 \textmu s combining the QEC decoding and error distribution time.

Overall, despite differing code distances and experimental settings, this comparison highlights that tightly integrating the decoder with the control and communication fabric---as in our design---is crucial for achieving sub-microsecond, end-to-end QEC feedback.


\section{Related Work}
\label{sec:related}

This section reviews prior work on qubit control systems, QEC decoding accelerators, and real-time QEC system implementations, and compares them with our integrated approach.
Our work presents the first implementation that integrates a scalable qubit control system with a low-latency QEC decoder.

\subsection{Qubit Control Systems}

As quantum hardware advances, increasingly sophisticated control requirements continue to drive active research into high-performance control architectures.
A variety of control systems have been developed to enable precise, pulse-level control of qubits across multiple hardware platforms.
Most existing systems are proprietary or closed-source~\cite{fu2017experimental,fu2019eqasm,xiang2020simultaneous,zhang2021exploiting,gebauer2023qicells,guo2023hisep,zhao2025distributed}, limiting their reproducibility and extensibility.
In contrast, several open-source frameworks provide publicly available designs, including ARTIQ\cite{kasprowicz2020artiq}, QICK\cite{stefanazzi2022qick}, QubiC\cite{xu2023qubic}, and RISC-Q\cite{liu2025risc}.
ARTIQ is based on custom hardware designed for AMO systems, whereas QICK, QubiC, and RISC-Q are implemented on AMD RFSoC platforms, offering greater accessibility and integration with superconducting qubit control.
Among these, RISC-Q provides a highly modular and parameterized generator for control systems, built using a high-level hardware description language without sacrificing performance.
This design flexibility makes RISC-Q particularly suitable for implementing scalable control architectures and integrating customized real-time feedback pipelines.
Accordingly, our QEC system is built upon the RISC-Q generator as the foundation for its qubit control subsystem.

\subsection{QEC Decoding Accelerators}

Although QEC decoding algorithms have been studied extensively for decades, real-time hardware implementations of QEC decoders have only recently become a focus of active research.
To achieve both low latency and high throughput, several hardware decoding accelerators have been developed \cite{das2022afs,das2022lilliput,vittal2023astrea,liao2023wit,liyanage2024fpga,barber2025real,wu2025micro,liyanage2025network,maan2025decoding,maurer2025real,zhang2025latte}.
These systems target different code families, exploring a variety of parallelization strategies and algorithmic optimizations.
Among the available implementations, Helios\cite{liyanage2024fpga} and Micro Blossom\cite{wu2025micro} stand out as the only open-source hardware decoders to date.
Helios implements a fully hardware-based union-find decoder optimized for low and stable latency, while Micro Blossom adopts a hybrid design in which part of the decoding logic runs on a CPU for higher accuracy.
Because our system emphasizes predictable real-time performance and tight hardware coupling, we integrate Helios as the decoder in our prototype implementation.

\section{Conclusion}
We present a scalable, open-source QEC system with sub-microsecond decoding–feedback latency and evaluate its performance. Estimates based on the evaluation data indicate that the decoding–feedback latency can be maintained in the sub-microsecond regime as the system scales to 881 qubits, confirming the feasibility of implementing a full decode–feedback loop within Google’s 1.1~\textmu s QEC cycle time. These results suggest that our architecture provides a strong foundation for large-scale, fault-tolerant quantum computing. Future work includes extending this approach to a complete fault-tolerant quantum computing stack in which Pauli frames are integrated to support real-time logical measurements and non-Clifford gate operations, as well as adapting the system to quantum platforms beyond superconducting qubits.

\section*{Acknowledgement}
We are grateful to Yue Wu, Lin Zhong, Namitha Liyanage, Yingkang Cao, Mark Horowitz, and Jason Cong for helpful discussions. 
This project is supported by Air Force Office of Scientific Research under award number FA9550-21-1-0209, the U.S. National Science Foundation grant CCF-1942837 (CAREER), CCF-2330974, NQVL-2435244, the U.S. Department of Energy, Office of Science, Advanced Scientific Computing Research
Testbeds for Science program, the National Quantum Information Science Research Centers Quantum Systems Accelerator, and a Sloan Research Fellowship.

\balance
\bibliographystyle{IEEEtran}
\bibliography{riscq}

\end{document}